\documentclass[letterpaper,twocolumn,showpacs,pra,aps]{revtex4-1}

\usepackage{amsmath}  
\usepackage{amsfonts} 
\usepackage{graphicx} 
\usepackage{amssymb}

\begin{document}

\title{Natural analog to cosmology in basic condensed matter physics}

\author{Eugene B. Kolomeisky}

\affiliation
{Department of Physics, University of Virginia, P. O. Box 400714, Charlottesville, Virginia 22904-4714, USA}

\date{\today}

\begin{abstract}
We show that the spatially homogeneous and isotropic evolution of a macroscopic Coulomb system of identical particles obeys equations that have the structure of the cosmological equations of the general theory of relativity.  There is a Hubble law, and the background charge (if present) mimics the effect of a negative cosmological constant.  Specifically, Coulomb explosions mimic the non-singular open cosmologies in negatively curved spaces, while breathing modes in conductors model oscillatory universes including the anti-de Sitter space.              
\end{abstract}


\maketitle

\section{Introduction}

Jellium -- a one-component plasma of interacting electron gas in the presence of a uniform positive background charge (representing the effect of ions) \cite{Wigner}-- is a paradigm of physics.  In condensed matter systems it has shaped our understanding of metals and doped semiconductors \cite{Bohm_Pines,Pines_Nozieres,Mahan}.  The Jellium model also successfully describes the interior of white dwarfs \cite{Salpeter} where the roles of the electrons and ions are interchanged:  nuclei move in a uniform gas of electrons.  In what follows we will use the condensed matter description (moving electrons) but the conclusions hold for any system of identical charged particles in the presence of an oppositely charged background.

The long-wavelength properties of Jellium can be derived from a classical macroscopic theory that combines hydrodynamics and electrodynamics \cite{Feynman,Jackson}.  In this approach the electrons are treated as an ideal charged liquid characterized by the local position- and time-dependent number density $n(\textbf{r},t)$ and velocity $\textbf{v}(\textbf{r},t)$ fields.  These are related by the continuity equation
\begin{equation}
\label{continuity}
\frac{\partial n}{\partial t}+\nabla \cdot(n\textbf{v})=0.
\end{equation}
The equation of motion of the liquid is given by the Euler equation of hydrodynamics 
\begin{equation}
\label{2nd_law}
\frac{\partial \textbf{v}}{\partial t} +(\textbf{v}\cdot \nabla)\textbf{v}= \frac{e}{m}\left (\textbf{E}+\frac{1}{c}\textbf{v}\times \textbf{B}\right )-\frac{1}{mn}\nabla p
\end{equation}
where $m$ is the electron mass, $e$ is its charge, $\textbf{E}$ is the electric field, $\textbf{B}$ is the magnetic field, $c$ is the speed of light, and $p$ is the pressure.  The electron liquid is accelerated both by the Lorentz force, the first term in the right-hand side in Eq.(\ref{2nd_law}), and by the gradient of the pressure $\nabla p$.  The electric and magnetic fields in turn are determined by Maxwell's equations \cite{Jackson}, specifically, Gauss's law of electricity,
\begin{equation}
\label{electric_Gauss}
\nabla\cdot\textbf{E}=4\pi e(n-n_{0}),
\end{equation}
where $n_{0}$ is the number density of the positively charged background,  Gauss's law of magnetism,
\begin{equation}
\label{magnetic _Gauss}
\nabla\cdot \textbf{B}=0,
\end{equation} 
Faraday's law,
\begin{equation}
\label{Faraday}
\nabla\times\textbf{E}=-\frac{1}{c}\frac{\partial \textbf{B}}{\partial t},
\end{equation}
and Amp\`ere-Maxwell's law,
\begin{equation}
\label{AM}
\nabla\times \textbf{B}=\frac{1}{c}\frac{\partial \textbf{E}}{\partial t}+\frac{4\pi}{c}en\textbf{v}.
\end{equation}
The charge density is $\rho=e(n-n_{0})$ while the current density is $\textbf{j}=en\textbf{v}$. 

While for an equilibrium  sample of net zero charge the local neutrality $n=n_{0}$ holds, a small uniform displacement of the electron liquid relative to the background triggers a harmonic oscillation with the plasma frequency $\omega_{p}$  given by \cite{Feynman,Jackson}
\begin{equation}
\label{plasma_frequency}
\omega_{p}^{2}=\frac{4\pi n_{0}e^{2}}{m}
\end{equation}    

The goal of this work is to show that Jellium can support a novel uniform and isotropic nonlinear density oscillation that in a special case reduces to the uniform plasma oscillation with the frequency $\omega_{p}$ (\ref{plasma_frequency}).  Surprisingly, the dynamics is governed by equations that have the same structure as the Friedmann-Lema\^{\i}tre-Robertson-Walker (FLRW) cosmological equations of the general theory of relativity in the presence of the cosmological constant \cite{ZN}.  The condensed matter model is antipodal to the FLRW cosmology in that the attractive gravitational interaction is replaced by the repulsive Coulomb interaction of the electrons, while the attractive interaction between the electrons and the background mimics the effect of a \textit{negative} cosmological constant.  

Incidentally, for over two decades theorists have been captivated by the anti-de Sitter/conformal field theory (AdS/CFT) correspondence conjecture \cite{AdS} that implies that fundamental particle interactions may be described geometrically in terms of the AdS space.   The latter is specified by a negative cosmological constant and lack of matter.  The AdS space is an antipode of the de Sitter (dS) space characterized by a positive cosmological constant;  in its current stage of expansion the Universe is approximately a dS space \cite{SS}.   The AdS/CFT correspondence also provides tools to tackle problems in several other areas of theoretical physics \cite{book1}.  Particularly, in condensed matter theory the AdS/CFT techniques have been applied to systems without quasiparticles \cite{book2}.  

Below we demonstrate (without resorting to the AdS/CFT ideas) that due to the presence of charged background an AdS space can be realized in Jellium.  

In the special case that the charged background is absent our findings are also relevant to the dynamics of a Coulomb explosion \cite{CE}.  This effect can be caused by an interaction of an intense ultrashort laser pulse with matter.  The large electric field of the pulse drives the light electrons away from the heavy atoms, leaving positively charged ions  behind; these undergo an explosion driven by the Coulomb repulsion.  This setting may be also viewed as an antipode to the FLRW cosmology without a cosmological constant.   

Our analysis is stimulated by a remarkable observation due to Milne and McCrea \cite{Milne,MM,McCrea0,McCrea1} who demonstrated that the FLRW equations, originally derived from Einstein's general theory of relativity, can be obtained within the framework of Newtonian gravity.  The Newtonian approach to cosmology was initiated by Mason \cite{Mason} who however did not look for uniform and isotropic solutions to the equations of motion.     

\section{Evolution equations} 

We seek solutions to Eqs.(\ref{continuity})-(\ref{AM}) corresponding to a spatially homogeneous and isotropic electron density $n=n(t)$.  Then at any instant $t$ the pressure $p$ (which is a function of $n$) is also uniform, so that $\nabla p=0$.  As a result, the second term in the right-hand side of the Euler equation (\ref{2nd_law}) is identically zero.  The space isotropy also dictates that in the rest frame of one of the electrons the remainder of the electron liquid is characterized by a radially symmetric velocity field, $\textbf{v}(\textbf{r},t)=v(r,t)\textbf{r}/r$, where the radius vector $\textbf{r}$ is the position relative to the electron at rest.  There can be no magnetic field, because it would have to be radially symmetric, which violates Gauss's law of magnetism (\ref{magnetic _Gauss}).  Then the dynamics of the electron liquid is governed by the electric field that follows from Gauss's law (\ref{electric_Gauss}):          
\begin{equation}
\label{electric_field}
\textbf{E}=\frac{4\pi e}{3}[n(t)-n_{0}]\textbf{r}.
\end{equation}
With all this in mind, the continuity (\ref{continuity}) and the Euler (\ref{2nd_law}) equations can be written as
\begin{equation}
\label{radial_continuity}
\frac{\dot{n}}{n}+\frac{(r^{2}v)'}{r^{2}}=0
\end{equation} 
\begin{equation}
\label{radial_Euler}
\frac{\dot{v}+vv'}{r}=\frac{4\pi e^{2}}{3m}(n-n_{0})
\end{equation}
where the dot and the prime are shorthands for the derivatives with respect to $t$ and $r$.  Below we follow McCrea and Milne \cite{MM}, modifying their reasoning appropriately to the problem at hand.

First, let us introduce a new function $H(t)$ such as
\begin{equation}
\label{Hubble_definition}
\dot{n}=-3H(t)n
\end{equation}
Then the continuity equation (\ref{radial_continuity}) becomes 
\begin{equation}
\label{radial_continuity1}
(r^{2}v)'=3H(t)r^{2} 
\end{equation}
and can be integrated with the result
\begin{equation}
\label{v(r,t)}
v=H(t)r+\frac{F(t)}{r^{2}}
\end{equation}
Since the right-hand side in Eq.(\ref{radial_Euler}) only depends on $t$, the same must be true for its left-hand side. The velocity field (\ref{v(r,t)}) satisfies this requirement if $F(t)\equiv 0$.  Restoring the vector notation, Eq.(\ref{v(r,t)}) becomes
\begin{equation}
\label{Hubble_law}
\textbf{v}=H(t)\textbf{r}
\end{equation}
which looks like the Hubble law with $H(t)$ being the Hubble parameter \cite{ZN}.  We now observe that Eqs.(\ref{electric_field}), (\ref{Hubble_definition}) and (\ref{Hubble_law}) automatically satisfy Faraday's and Amp\`ere-Maxwell's laws, Eqs.(\ref{Faraday}) and (\ref{AM}), with $\textbf{B}=0$.

Substituting Eq.(\ref{Hubble_law}) into the Euler equation (\ref{radial_Euler}) we find 
\begin{equation}
\label{Friedmann1}
\dot{H}+H^{2}(t)=\frac{4\pi e^{2}}{3m}(n-n_{0})
\end{equation}
hereafter called the Coulomb-Newton equation.

The evolution described by Eqs.(\ref{Hubble_definition}), (\ref{Hubble_law}) and (\ref{Friedmann1}) refers to the rest reference frame of one of the electrons:  the remainder of the electron liquid moves radially relative to the chosen electron.  This does not imply that there is a preferred origin $O$ associated with the electron at rest \cite{ZN,McCrea1}.  Indeed, let us attach another reference frame to an electron of position vector $\mathbf{R}$ at $O'$ that is moving relative to $O$ with velocity $\mathbf{V}$. The radius vectors $\mathbf{ r}$ and $\mathbf{r}'$ of an arbitrary point of the liquid, and its velocity vectors $\mathbf{v}$ and $\mathbf{v}'$ in the two reference frames are related by a Euclidian transformation for the positions and a Galilean transformation for the velocities:
\begin{equation}
\label{transformation}
\mathbf{r}=\mathbf{r}'+\mathbf{R}, ~~~\mathbf{v}=\mathbf{v}'+\mathbf{V}.
\end{equation}  
Inserting Hubble's law (\ref{Hubble_law}), $\mathbf{ v}=H\mathbf{ r}$ and $\mathbf{ V}=H\mathbf{ R}$ into the second of Eqs.(\ref{transformation}) we find $\mathbf{v}'=H(\mathbf{r}-\mathbf{ R})=H\mathbf{ r}'$ which is again the same Hubble's law (\ref{Hubble_law}).  So the motion from the viewpoint of an observer at point $O'$ (and moves with it) is exactly the same as that from the viewpoint of an observer at point $O$.  Therefore (neglecting finite-size effects) the solution contained in Eqs.(\ref{Hubble_definition}), (\ref{Hubble_law}) and (\ref{Friedmann1}), satisfies the requirement of homogeneity and isotropy.  The Hubble law (\ref{Hubble_law}) is then a statement about the relative velocity $\mathbf{v}$ of two particles separated by the vector $\mathbf{r}$.

Eq.(\ref{Hubble_definition}) has the same form as the corresponding cosmological equation (also known as the second Friedmann equation) in the matter-dominated regime \cite{MM,ZN}, while the Coulomb-Newton equation (\ref{Friedmann1}) only differs from its matter-dominated counterpart in the overall sign in the right-hand side (as demonstrated below).  The Hubble law (\ref{Hubble_law}) looks exactly the same both in the Coulomb and Newtonian cases.    To appreciate similarities and differences between the two theories we now outline the Newtonian cosmology of Milne and McCrea \cite{Milne,MM,McCrea0,McCrea1,ZN}.  

\section{Newtonian cosmology} 

Let us consider non-relativistic gravitating matter of particles of mass $m$.  As in the Jellium case, a hydrodynamic description in terms of the number density $n$ and velocity $\mathbf{ v}$ fields is adopted.  Then the conservation of mass is expressed by the same equation as before (\ref{continuity}).  The corresponding Euler equation has the same form
\begin{equation}
\label{gravitational_Euler}
\frac{\partial \textbf{v}}{\partial t} +(\textbf{v}\cdot \nabla)\textbf{v}= \mathbf{g}-\frac{1}{mn}\nabla p
\end{equation} 
as the $\mathbf{B}=0$ limit of Eq.(\ref{2nd_law}).  The gravitational field $\mathbf{g}$ is determined by Gauss's law of gravity \cite{MM,Bonnor,ZN}
\begin{equation}
\label{gravitational_Gauss}
\nabla\cdot\textbf{g}=-4\pi Gm(n-n_{0}),~~~n_{0}=\frac{\Lambda c^{2}}{4\pi G m}
\end{equation}
where $G$ is the universal gravitational constant and the characteristic number density $n_{0}$, a counterpart of the number density of the background charge in Eq.(\ref{electric_Gauss}), is finite only in the presence of Einstein's cosmological constant $\Lambda$.  This equation differs from Gauss's law (\ref{electric_Gauss}) in the overall sign in the right-hand side.  Comparing the two Euler equations (\ref{2nd_law}) and (\ref{gravitational_Euler}) and the two Gauss's laws (\ref{electric_Gauss}) and (\ref{gravitational_Gauss}) the Coulomb-Newton correspondence can be summarized in a relationship
\begin{equation}
\label{correspondence}
e^{2}\rightarrow -Gm^{2}.
\end{equation}

Gauss's law of gravity (\ref{gravitational_Gauss}) encompasses physics which is qualitatively different from that of Jellium:  particles attracted to each other gravitationally are being repelled by a background due to the cosmological constant.  The most telling manifestation of this difference is the fate of the $n=n_{0}$ solution to Eqs.(\ref{gravitational_Euler}) and (\ref{gravitational_Gauss}) which corresponds to Einstein's static Universe.   This solution is unstable \cite{Bonnor} which can be also seen from an expression for gravitational counterpart $\omega_{G}$ of the plasma frequency (\ref{plasma_frequency}) that can be recovered with the help of Eq.(\ref{correspondence}),
\begin{equation}
\label{Jeans_frequency}
\omega_{G}^{2}=-4\pi Gmn_{0}=-\Lambda c^{2}<0.
\end{equation}  
On the other hand, the $n=n_{0}$ solution of the Coulomb case corresponding to the state of local neutrality is stable since $\omega_{p}^{2}>0$, Eq.(\ref{plasma_frequency}).

Finally, applying the substitution (\ref{correspondence}) to Eq.(\ref{Friedmann1}) we obtain an equation
\begin{equation}
\label{Friedmann2}
\dot{H}+H^{2}(t)=-\frac{4\pi Gm}{3}(n-n_{0})
\end{equation}
known as the Newton equation for the Universe which along with Eqs.(\ref{Hubble_definition}) and (\ref{Hubble_law}) summarizes the FLRW cosmology in the presence of the cosmological constant \cite{MM,ZN} in the matter-dominated regime.  The Newton equation for the Universe (\ref{Friedmann2}) only differs from the Coulomb-Newton equation (\ref{Friedmann1}) in the overall sign in the right-hand side.

We also note that if $p=0$ the FLRW cosmological equations can be derived from the point particle perspective that avoids continuum treatment \cite{EG}.  For the same discrete cosmological model initial conditions leading to the FLRW-type evolution have been also discussed \cite{BGS}.   

Let us introduce a new function $a(t)$ called the scale factor \cite{ZN} such as 
\begin{equation}
\label{scale_factor}
H(t)=\frac{\dot{a}}{a}.
\end{equation}
In the FLRW cosmology $a(t)$ describes expansion or contraction of space itself.  Indeed, substituting this expression into Hubble's law (\ref{Hubble_law}), employing $\textbf{v}=d\textbf{r}/dt$, and integrating following the motion we find
\begin{equation}
\label{comoving}
\textbf{r}=a(t)\textbf{x}
\end{equation}
where $\textbf{x}$ is a time-independent comoving coordinate vector defining the particle considered.  Vector $\textbf{x}$ factors out the dynamics of the space fabric and the scale factor $a(t)$ relates it to the proper (laboratory) position of the particle $\textbf{r}$.  On the other hand, Milne and McCrea adopt a complementary point of view that space is static and the world fluid is either expanding or contracting.   

Combining Eqs.(\ref{scale_factor}) and (\ref{Hubble_definition}), the latter can be integrated with the result 
\begin{equation}
\label{charge_conservation}
n(t)=\frac{\beta}{a^{3}(t)}
\end{equation}
(where $\beta$ is an integration constant) which is a statement of conservation of the number of particles within evolving sphere of radius $a(t)$.

Substituting Eqs.(\ref{scale_factor}) and (\ref{charge_conservation}) into the Newton equation for the Universe (\ref{Friedmann2}), we find an expression
\begin{eqnarray}
\label{scale_factor_equation0}
m\ddot{a}&=&-\frac{4\pi Gm^{2}(n-n_{0})a^{3}}{3a^{2}}\nonumber\\
&=&-\frac{4\pi Gm^{2}}{3}\left (\frac{\beta}{a^{2}}-n_{0}a\right )\equiv F(a)
\end{eqnarray}
that is the equation of motion for a particle on the surface of an expanding or contracting sphere of radius $a$ caused by the force $F(a)$.   The energy integral of Eq.(\ref{scale_factor_equation0}) has the form 
\begin{equation}
\label{energy_integral0}
\frac{m\dot{a}^{2}}{2}+U(a)=E
\end{equation}
where
\begin{equation}
\label{potential_energy0}
U(a)=-\int F(a)da=-\frac{4\pi Gm^{2}}{3}\left (\frac{\beta}{a}+\frac{n_{0}a^{2}}{2}\right )
\end{equation}
is the potential energy and $E$ is the energy.  The conservation laws (\ref{charge_conservation}), (\ref{energy_integral0}) and (\ref{potential_energy0}) can be combined into an expression \cite{ZN,MM,McCrea0,Bonnor}
\begin{equation}
\label{1stFriedmann0}
\left (\frac{\dot{a}}{a}\right )^{2}=\frac{8\pi Gm}{3}\left (n+\frac{n_{0}}{2}\right )-\frac{kc^{2}}{a^{2}}, ~~~k=-\frac{2E}{mc^{2}}
\end{equation}  
that is also called the first Friedmann equation.

In the cosmology the scale factor $a(t)$ is the radius of space curvature while the combination $k/a^{2}$ in the last term in (\ref{1stFriedmann0}) is (loosely) called the Gaussian curvature of space:  the $k>0$ space possesses positive curvature and is closed, the $k<0$ space is negatively curved and open, while $k=0$ describes Euclidian (flat) space.      

The dynamics of the scale factor $a(t)$ accumulated in Eq. (\ref{energy_integral0}) can be understood in terms of the classical motion of a particle of mass $m$ and energy $E$ in the field of the potential energy $U(a)$ (\ref{potential_energy0}).  While the possible outcomes are well-documented \cite{ZN,MM,McCrea0,Bonnor}, one class of solutions is of particular interest to the present study:

In the absence of the cosmological constant, $n_{0}=0$, the potential energy (\ref{potential_energy0}) becomes $U(a)=-4\pi \beta Gm^{2}/3a$, and the resulting cosmologies can be inferred from the solution to the zero angular momentum \textit{attractive} Kepler's problem \cite{LL1}.   Here the three well-known types of motion, elliptic, parabolic, and hyperbolic, correspond to closed ($E<0$ or $k>0$), flat ($E=0$ or $k=0$), or open ($E>0$, $k<0$) cosmologies, respectively \cite{Milne,MM,McCrea0,McCrea1}.  

Finally, the first Friedmann equation (\ref{1stFriedmann0}) can be presented in an equivalent form that only involves measurable quantities, specifically,  Hubble's parameter (\ref{scale_factor}) and the density (\ref{charge_conservation}) 
\begin{equation}
\label{Friedmann3}
H^{2}=\frac{8\pi Gm}{3}\left (n+\frac{n_{0}}{2}\right )+\gamma n^{2/3},\gamma=\frac{2E}{m\beta^{2/3}}=-\frac{kc^{2}}{\beta^{2/3}}
\end{equation}
where $\gamma$ is the only free parameter of the problem set by initial conditions \cite{MM,ZN}. 

\section{Solving the Coulomb-Newton evolution equations}

The Coulomb counterparts to Eqs.(\ref{scale_factor_equation0}), (\ref{potential_energy0}), (\ref{1stFriedmann0}), and (\ref{Friedmann3}) can be written out automatically with the help of the Coulomb-Newton correspondence relation (\ref{correspondence}):
\begin{eqnarray}
\label{scale_factor_equation1}
m\ddot{a}&=&\frac{4\pi e^{2}(n-n_{0})a^{3}}{3a^{2}}\nonumber\\
&=&\frac{4\pi e^{2}}{3}\left (\frac{\beta}{a^{2}}-n_{0}a\right )\equiv F(a),
\end{eqnarray}
\begin{equation}
\label{potential_energy}
U(a)=-\int F(a)da=\frac{4\pi e^{2}}{3}\left (\frac{\beta}{a}+\frac{n_{0}a^{2}}{2}\right ),
\end{equation}
\begin{equation}
\label{Coulomb_Friedmann}
\left (\frac{\dot{a}}{a}\right )^{2}=-\frac{8\pi e^{2}}{3m}\left (n+\frac{n_{0}}{2}\right )-\frac{kc^{2}}{a^{2}}, 
\end{equation}
\begin{equation}
\label{Friedmann4}
H^{2}=-\frac{8\pi e^{2}}{3m}\left (n+\frac{n_{0}}{2}\right )+\gamma n^{2/3}.
\end{equation}
Compared to their Newtonian analogs, Eqs.(\ref{scale_factor_equation0}), (\ref{potential_energy0}), (\ref{1stFriedmann0}), and (\ref{Friedmann3}), they all feature the sign change in the terms describing Coulomb interactions that has already been mentioned.   

The dynamics of the scale factor $a(t)$ accumulated in Eqs.(\ref{energy_integral0}) and (\ref{potential_energy}) is qualitatively different depending on whether the neutralizing background charge is present or not, and these two cases are analyzed separately.   

\subsection{Dynamics of a Coulomb explosion} 

When $n_{0}=0$, one sees from Eq.(\ref{potential_energy}) that $U(a)=4\pi e^{2}\beta/3a$, thus implying that we are essentially dealing with a zero angular momentum \textit{repulsive} Kepler's problem \cite{LL1}.  It has solutions only if $E>0$, and the motion is infinite.  Since $U(a\rightarrow \infty)\rightarrow 0$, there is a "terminal velocity"  $\dot{a}_{\infty}=\sqrt{2E/m}$.  This means that asymptotically the evolution is ballistic, $a(|t|\rightarrow\infty)=\sqrt{2E/m}|t|$, $H(|t|\rightarrow\infty)=\pm 1/|t|$, $n(|t|\rightarrow \infty)=1/\gamma^{3/2}t^{3}$, and the Coulomb effects are negligible. 

More generally, there is a turning point, $U(a)=E$, given by $a_{1}=4\pi e^{2}\beta/3E$ which corresponds to the smallest scale factor and thus to the largest electron density
\begin{equation}
\label{largest_density}
n_{1}=\frac{\beta}{a_{1}^{3}}=\left (\frac{3m \gamma}{8\pi e^{2}}\right )^{3}.
\end{equation}   
This is the density at which Hubble's parameter (\ref{Friedmann4}) vanishes, and the motion changes from contraction to expansion.  Existence of the density scale (\ref{largest_density}) allows us to introduce a corresponding time scale
\begin{equation}
\label{time_scale}
t_{1}=\left (\frac{3m}{8\pi n_{1}e^{2}}\right )^{1/2}=\frac{8\pi e^{2}}{3m\gamma^{3/2}}.
\end{equation}
Its physical significance consists in separating the ballistic regime ($|t|\gtrsim t_{1}$) from the Coulomb repulsion dominated regime ($|t|\lesssim t_{1}$)  of evolution (where the zero of time was chosen at $n=n_{1}$).  The density and time scales (\ref{largest_density}) and (\ref{time_scale}), respectively, define a convenient system of units hereafter adopted for the $n_{0}=0$ case.  Then Eq.(\ref{Friedmann4}) becomes
\begin{equation}
\label{H(n)_coulomb_explosion}
H^{2}=-n+n^{2/3}
\end{equation} 
We see that relative expansion or contraction rates are maximal at the density $n=8/27$.

\begin{figure}
\includegraphics[width=1.0\columnwidth, keepaspectratio]{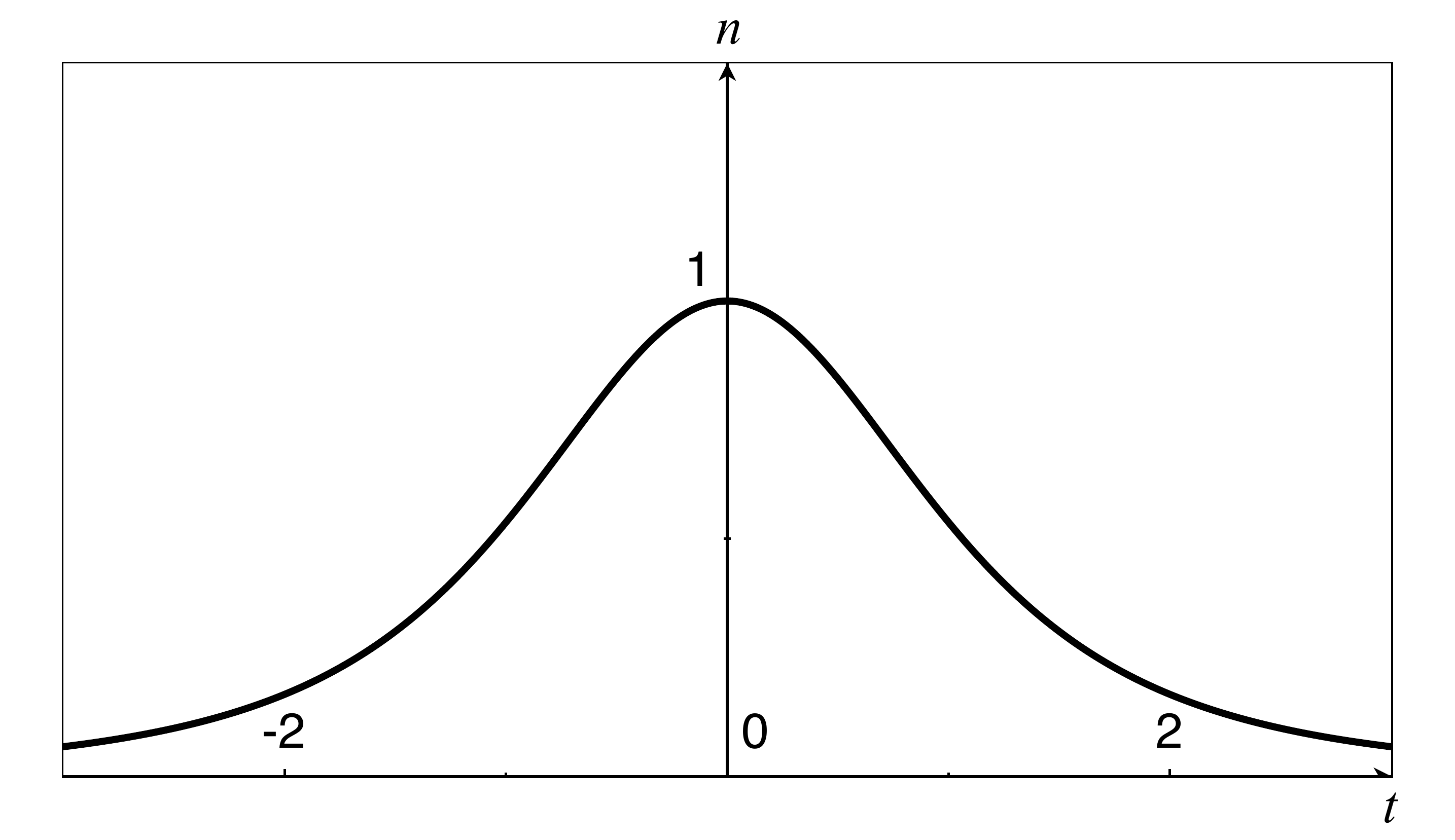} 
\caption{Evolution of the electron density $n(t)$ according to Eq.(\ref{densityvstime}) in units given by Eqs.(\ref{largest_density}) and (\ref{time_scale}) without positively charged background.  The electron liquid contracts for $t<0$ until it reaches the maximal density at $t=0$ followed by an expansion for $t>0$.}
\label{density_vs_time}
\end{figure}

Employing the solution $a(t)$ to the repulsive Kepler's problem \cite{LL1} along with Eq.(\ref{charge_conservation}), the time dependence of the density can be given in the following form
\begin{equation}
\label{densityvstime}
n=\frac{1}{a^{3}}=\frac{8}{(\cosh\xi+1)^{3}},~~~t=\frac{1}{2}(\sinh\xi+\xi)
\end{equation}  
where $\xi$ is a parameter varying from minus to plus infinity.  The dependence of the density on time (\ref{densityvstime}) is shown in Figure \ref{density_vs_time}.    

We observe that evolution of the electron liquid without a charged background features a universality encapsulated in Eqs.(\ref{H(n)_coulomb_explosion}) and (\ref{densityvstime}).    

Based on the form of the Coulomb counterpart (\ref{Coulomb_Friedmann}) of the first Friedmann equation (\ref{1stFriedmann0}) we conclude that the evolution of a Coulomb system of identical particles without an oppositely charged background is hyperbolic ($E>0$), the corresponding analog cosmology is open, and the analog space is negatively curved ($k<0$).  It differs qualitatively from the hyperbolic ($E>0$, $k<0$) FLRW cosmology  as the Big Bang type density singularity is prevented by the repulsive Coulomb interactions between the particles.   

\subsection{Breathing mode in Jellium} 

We now restore the original physical units and focus on the $n_{0}\neq0$ case, which means that the electron liquid evolves in the presence of an oppositely charged background of fixed number density $n_{0}$.   In this case the force $F(a)$ entering the right-hand side of Eq.(\ref{scale_factor_equation1}) vanishes at $a=a_{0}=(\beta/n_{0})^{1/3}$.  According to the particle conservation law (\ref{charge_conservation}) this happens when $n=n_{0}$, which is what is expected when neutrality holds locally.  The electron liquid is compressed if $a<a_{0}$ ($n>n_{0}$) and stretched if $a>a_{0}$ ($n<n_{0}$).  For small $a-a_{0}$, the force (\ref{scale_factor_equation1}) is linear in the difference, $F=-m\omega_{p}^{2}(a-a_{0})$, and restoring.  This is a harmonic oscillation with the plasma frequency $\omega_{p}$ (\ref{plasma_frequency}).  The existence of the characteristic density $n_{0}$, scale factor $a_{0}$ and time $1/\omega_{p}$ scales prompts adoption of the following units
\begin{equation}
\label{units}
[n]=n_{0},~~~~~ [a]=\left (\frac{\beta}{n_{0}}\right )^{1/3},~~~~~ [t]=\left (\frac{m}{4\pi n_{0}e^{2}}\right )^{1/2}
\end{equation}
where $[z]$ stands for the unit of $z$.  Then Eqs.(\ref{charge_conservation}), (\ref{energy_integral0}), (\ref{potential_energy}), and (\ref{Friedmann4}) acquire the form  
\begin{equation}
\label{charge_conservation2}
n(t)=\frac{1}{a^{3}(t)},
\end{equation} 
\begin{equation}
\label{energy_integral2}
\frac{\dot{a}^{2}}{2}+U(a)=\epsilon,~~ \epsilon=\frac{E}{4\pi e^{2}(n_{0}\beta^{2})^{1/3}}=\frac{m\gamma}{8\pi e^{2}n_{0}^{1/3}},
\end{equation}
\begin{equation}
\label{potential_energy2}
U(a)=\frac{1}{3} \left (\frac{1}{a}+\frac{a^{2}}{2}\right ),
\end{equation}
\begin{equation}
\label{H(n)_jellium}
H^{2}=-\frac{1}{3}(2n+1)+2\epsilon n^{2/3}.
\end{equation}

\begin{figure}
\includegraphics[width=1.0\columnwidth, keepaspectratio]{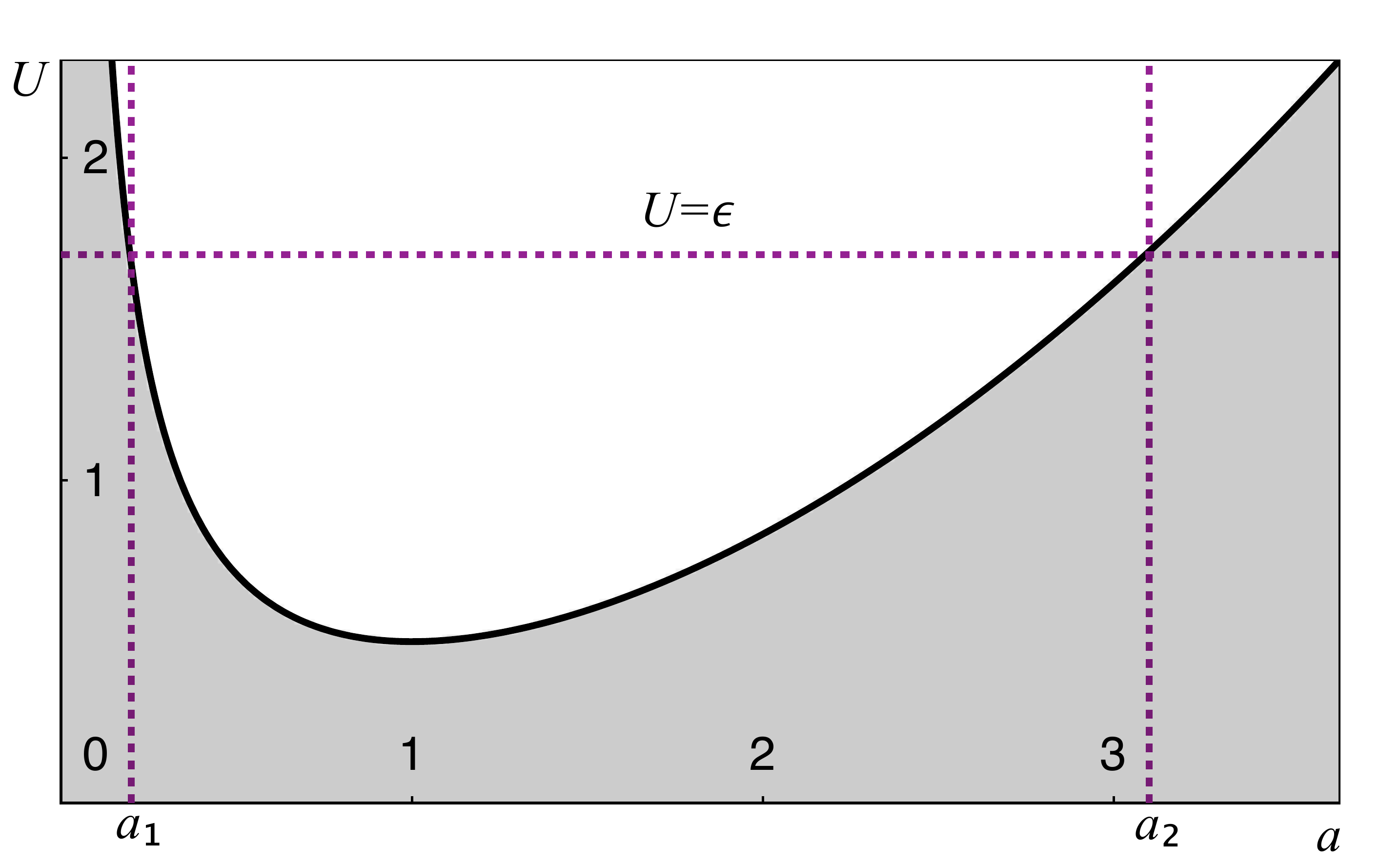} 
\caption{(Color online) Potential well $U(a)$ (\ref{potential_energy2}) for a particle of position $a$ and energy $\epsilon$ in units of Eq.(\ref{units}).  If $\epsilon\geqslant 1/2$ the motion is finite with turning points $a_{1,2}$ given by solutions to the equation $U(a)=\epsilon$.  Greyscale region is not accessible for the motion.}
\label{pwell}
\end{figure}

The dynamics of the scale factor $a(t)$ can be understood in terms of the motion of a classical particle of energy $\epsilon$ in the potential well (\ref{potential_energy2}) as sketched in Figure \ref{pwell}.  
Since $U(a)\geqslant 1/2$, Eq.(\ref{energy_integral2}) has solutions only if $\epsilon\geqslant 1/2$.  When $\epsilon=1/2+0$ the scale factor undergoes simple harmonic oscillation as already described with unit (plasma) frequency while for $\epsilon>1/2$ the oscillation is nonlinear and asymmetric.  For $\epsilon$ fixed, respectively, the turning points $a_{1}$ and $a_{2}$ are the smallest and largest solutions to the cubic equation $U(a)=\epsilon$.  Correspondingly, the electron density oscillates between its largest and smallest values, $n_{1}=1/a_{1}^{3}$ and $n_{2}=1/a_{2}^{3}$, which are also zeros of the Hubble parameter (\ref{H(n)_jellium}).    This oscillation can be classified as a homogeneous monopole (breathing) mode.    

The dependence of the Hubble parameter on the density is given by Eq.(\ref{H(n)_jellium}):  expansion ($H>0$) and contraction ($H<0$) of the electron liquid take place within the density range $[n_{2},n_{1}]$ corresponding to the $[a_{1},a_{2}]$ interval of the scale factor of Figure \ref{pwell}.  The relative expansion or contraction rate is maximal at a density $n=(2\epsilon)^{3}$.  The density $n$ as a function of time $t$ can be obtained by integrating Eq.(\ref{energy_integral2}) 
\begin{equation}
\label{t_of_n}
t=\frac{1}{\sqrt{2}}\int_{a_{1}}^{n^{-1/3}}\frac{da}{\sqrt{\epsilon-U(a)}}
\end{equation}
while the period of the Jellium oscillation is given by
\begin{equation}
\label{period}
T=\sqrt{2}\int_{a_{1}}^{a_{2}}\frac{da}{\sqrt{\epsilon-U(a)}}
\end{equation}  
With potential energy function $U(a)$ as in Eq.(\ref{potential_energy2}) these expressions can be reduced to elliptic integrals.   

We conclude that the evolution of the Coulomb system of identical particles in the presence of an oppositely charged background gives an example of a non-singular oscillatory analog cosmology.

\subsection{Anti-de Sitter limit}

When the dimensionless energy parameter $\epsilon$ that appeared in Eqs.(\ref{energy_integral2}), (\ref{H(n)_jellium}), (\ref{t_of_n}), and (\ref{period}) is very large, the breathing mode approximately corresponds to the AdS cosmology.  Indeed, let us restore the original physical units and set the parameter $\beta$ appearing in Eqs.(\ref{charge_conservation}), (\ref{scale_factor_equation1}), and (\ref{potential_energy}) at zero.  This is the AdS space:  the electron density is zero, and the scale factor $a(t)$ undergoes a motion in the harmonic potential $U(a)=m\omega_{p}^{2}a^{2}/6$ interrupted by impenetrable wall at $a=0$:
\begin{equation}
\label{AdS}
a(t)=\left (\frac{6E}{m\omega_{p}^{2}}\right )^{1/2}\left |\sin\frac{\omega_{p}t}{\sqrt{3}}\right |
\end{equation}  
This infinite periodic sequence of bounces with collapses $a=0$ at integer multiples of the period of motion $T=\pi \sqrt{3}/\omega_{p}$ is unobservable until test charges are inserted into the system and their motion is monitored.  The charges in turn modify the dynamics but Eq.(\ref{AdS}) remains approximately correct, provided that $a(t)\gg a_{0}=(\beta/n_{0})^{1/3}$, where the latter factor is the position of the minimum of the potential energy (\ref{potential_energy}).  This is equivalent to the requirement $\epsilon\gg 1$, as also apparent from Figure \ref{pwell}.  Even in this case, due to Coulomb repulsions between the charges, Eq.(\ref{AdS}) breaks down in the vicinity of the zeros of $a(t)$ thus eliminating the collapses.  The width of these regions,  estimated as $\triangle t \simeq1/\omega_{p}\sqrt{\epsilon}$, is parametrically small compared to the oscillation period.  This means that over a significant portion of the Jellium oscillation the dynamics is dominated by the charged background.

In quantum mechanics the AdS nature of the classical $\epsilon\gg1$ breathing mode manifests itself in the character of the energy spectrum of highly-excited states in the potential well (\ref{potential_energy}):
\begin{equation}
\label{spectrum}
E_{l}\approx \frac{\hbar \omega_{p}}{\sqrt{3}}l
\end{equation}  
where $l\gg 1$ is odd.   Thus the energy of the AdS plasmon is $2\hbar \omega_{p}/\sqrt{3}$ as opposed to $\hbar \omega_{p}$ for the regular bulk plasmon.  It may be possible to excite AdS plasmons by passing charged particles through the material, and then the experimental signature of the AdS plasmons would be an energy loss equal to integral multiples of $2\hbar \omega_{p}/\sqrt{3}$.            

\section{Conclusions}  

To summarize, we have demonstrated that the uniform and isotropic evolution of a macroscopic Coulomb system of identical particles with or without an oppositely charged background obeys equations that have the form of the FLRW cosmological equations.  The corresponding Coulomb universes differ from the physical Universe as the role of the attractive gravity is played by the repulsive Coulomb interaction between the particles.  

In the heart of this analogy lies Newton's shell theorem \cite{shell} which states that gravitational field outside of a thin uniform sphere is the same as that of a particle at the center, having the same mass as the sphere, and the field is zero inside the sphere.  The shell theorem also holds for the Coulomb force law.  Our Eq.(\ref{electric_field}), as well as its gravitational cousin ($\textbf{g}\propto \textbf{r}$) may be viewed as consequences of the shell theorem.   Since a $d$-dimensional counterpart of Eq.(\ref{electric_field}) also has the $\textbf{E}\propto \textbf{r}$ form, solutions of this type exist in general dimension.  This however \textit{excludes} laboratory low-dimensional systems such as wires or layers imbedded in three-dimensional space.   Indeed, the electric field driving the evolution in these systems is only a projection of the total three-dimensional electric field onto the $d$-dimensional space of the Coulomb system.  On the contrary, Gauss's law or equivalently the shell theorem involve all components of the electric field.    

The hallmark of the FLRW-type evolution both in the Newtonian and Coulomb cases is the absence of a preferred origin:  every particle can be chosen as an origin.  However in practice a preferred origin can be imposed by finite-size effects.  For example, one can imagine a finite-radius uniformly charged ball of identical particles contracting or expanding into an unbounded charged background (or without any background at all).  The center of the ball is the preferred origin relative to which the evolution of the system is spherically-symmetric.  Our analysis covers this situation too, \textit{provided} $p=0$.  Indeed, the statement of charge conservation (\ref{charge_conservation}) and the equation of motion of a test particle (\ref{scale_factor_equation1}) on the surface of an expanding or contracting sphere of radius $a$ still apply to all the particles within the ball, including those on its physical boundary.  Choosing the integration constant in Eq.(\ref{charge_conservation}) to be $\beta=3N/4\pi$ (where $N$ is the total number of particles within the ball), the scale factor becomes the radius of the ball; its evolution $a(t)$ is also predicted by our theory.  For example, in the Coulomb explosion ($n_{0}=0$) case the radius of the ball evolves with time according to
\begin{equation}
\label{explosion}
a=\frac{a_{1}}{2}(\cosh\xi+1),~~~~~t=\frac{1}{2}\left (\frac{ma_{1}^{3}}{2e^{2}N}\right )^{1/2}(\sinh\xi+\xi)
\end{equation}                   

We hope that in the near future our results will be testable in Coulomb explosion experiments as well as in studies of nonlinear monopole modes.     

\section{Acknowledgements}  

I thank G. W. Gibbons for informing me of Refs.\cite{Mason}, \cite{EG} and \cite{BGS}, R. R. Jones for a discussion regarding Coulomb explosions, and J. P. Straley for valuable comments.

\end{document}